\documentstyle[aps,preprint,psfig]{revtex}
\tighten
\begin{document}

\draft
\title{Possible solution of the Coriolis attenuation problem}
\author{Pavlos Protopapas
\thanks{pavlos@walet.physics.upenn.edu}
and Abraham Klein
\thanks{aklein@walet.physics.upenn.edu}} 
\address{Department of Physics, University of Pennsylvania, Philadelphia,
PA 19104-6396}  

\date{\today}

\maketitle

\begin{abstract}
The most consistently useful simple model for the study of odd deformed
nuclei, the particle-rotor model (strong coupling limit of the 
core-particle coupling model) has nevertheless been beset by a long-standing
problem:  It is necessary in many cases to introduce an {\it ad hoc}
parameter that reduces the size of the Coriolis interaction coupling
the collective and single-particle motions.  Of the numerous suggestions
put forward for the origin of this supplementary interaction, none of 
those actually tested by calculations has been accepted as the solution
of the problem.  In this paper we seek a solution of the difficulty
within the framework of a general formalism that starts from the spherical 
shell model and is capable of treating an arbitrary linear combination
of multipole and pairing forces.  With the restriction of the interaction
to the familiar sum of a quadrupole multipole force and a monopole pairing 
force, we have previously studied 
a semi-microscopic version of the formalism whose framework is nevertheless
more comprehensive than any previously applied to the problem.  We obtained
solutions for low-lying bands of several strongly deformed odd
rare earth nuclei and found
good agreement with experiment, except for an exaggerated staggering of
levels for $K=\frac{1}{2}$ bands, which can be understood as a manifestation  
of the Coriolis attenuation problem.  We argue that within the formalism
utilized, the only way to improve the physics is to add interactions
to the model Hamiltonian.  We verify that by adding
a magnetic dipole interaction of essentially fixed strength, 
we can fit the $K=\frac{1}{2}$ bands without 
destroying the agreement with other bands.  
In addition we show that our solution
also fits $^{163}$Er, a classic test case of Coriolis attenuation 
that we had not previously studied.  
\end{abstract}

\begin{center}
PACS number(s): 21.60.-n, 21.60.Ev, 21.10.-k, 21.10.Re
\end{center}

\section{Introduction}

In this paper we propose a solution to a long-standing conundrum in the 
theoretical description of the properties of low-lying bands of 
deformed odd nuclei
within the framework of the particle-rotor model \cite{RS}.  Generally
speaking, the Hamiltonian of this model has the form
\begin{equation}
H= H_{\rm rotor} + H_{\rm particle}.  \label{ham1}
\end{equation}
Here the first term is a phenomenological operator describing the included
bands of a given even (core) nucleus, assumed to be a rotor.  
In the simplest case, it describes only
the ground band of an axially symmetric system.  
The second term
contains the single-particle energies of and the residual interactions
among the particles that move outside the rotor and 
are assumed to be kinematically independent of
the constituents of this core.  When we wish to describe odd nuclei
that are adjacent to the given core, the simplest assumption is that it 
suffices to consider
a single extra particle, or if we include pairing interactions, quasiparticle, 
outside the core,
described in the intrinsic system of the rotor by a set of axially
symmetric solutions of the Nilsson-BCS formalism.

An important talking point in favor of this extreme model is that it contains,
in so far as its predictions of energy levels is concerned, essentially 
no free parameters, since the quasiparticle properties can be determined
by fitting bandheads alone, and the core Hamiltonian depends to a good
approximation only on the observed moment of inertia of its ground-state
band.  When
the angular momentum of the core is replaced, according to angular
momentum conservation, by the difference between the total
angular momentum ${\bf J}$ and the particle angular momentum ${\bf j}$, 
the interaction between particle and core becomes (omitting BCS
occupation factors)
\begin{equation}
\frac{{\bf j}\cdot{\bf J}}{2{\cal I}}, \label{coriolis}
\end{equation}
where ${\cal I}$ is the moment of inertia of the core.  

For many applications, the model as described above does not predict the  
observed band structure \cite{DF5,DF6} unless
one multiplies the Coriolis interaction
by a factor $\xi\cong .6-.8$, thus giving
rise to the name Coriolis Attenuation.  This means that the model as described
above is too rudimentary, and that there is missing physics.
There have been numerous suggestions concerning the source of this 
missing physics.  We shall not review these suggestions in the form that 
they are described in the literature \cite{DF5,DF6} for at least
two reasons.  The first is that all those that have actually been 
tested by calculations fall short in some respect.
The second is that they are all stated within the
framework of some extended version of the particle-rotor model.  
But this model
has among its flaws the assumption of the kinematic independence
of the particle and rotor variables.  This certainly violates the Pauli
principle, the more so if one wishes to allow the particle to have access
to several major shells (as in the calculations we report in the present
work).

Instead we shall approach the subject from the point of view
of a theoretical framework that is formally exact and contains 
the particle-rotor model as a well-defined limit.  
We claim that for a sufficiently simple Hamiltonian we can, within this 
framework, compute
accurate solutions for the low lying bands of well-deformed
odd nuclei.  Therefore with high probability a failure of the theory
is to be ascribed to a missing term in the Hamiltonian. For example,
one of the classes of proposed explanation within the particle-rotor
framework, that we need more than one
particle outside the core, is ruled out in our case by the fact that the
exact formalism never involves more than one particle outside a band
of the core, the point being that the inclusion of core excitation
is taken care of by the inclusion of excited bands of the core.  Another
explanation of the standard list is that the model in its simplest form 
omits or treats incorrectly the 
so-called recoil term that arises from the same transformation 
that yields the Coriolis coupling; the distinction between the two terms  
is a consequence of the use of an intrinsic frame in the implementation
of the model.  But our calculations are carried
in the so-called laboratory system, in a way that automatically includes 
the effects of the recoil term.
The last important class of suggestions
and the only one that is cogent from the point of view 
of the theory we utilize is that of a missing interaction.  Here,
the structure of our equations suggests an essentially unique choice,
at least as far as multipolarity is concerned.  Further discussion
of this point, the major one made in this paper, is continued below, 
after some remarks about the formalism.

The theory of collective motion for odd nuclei that we shall apply 
is one that we have recently resuscitated and extended, and that 
we have rechristened the 
Kerman-Klein-D\"onau-Frauendorf (KKDF) model 
\cite{pav:1,pav:2,pav:3,pav:4,pav:5}.  
This model was introduced by D\"onau and
Frauendorf \cite{DF1,DF2,DF3,DF4,DF5}, whose work was in turn stimulated 
by an application
\cite{Dreiss} of the theory of collective motion developed by Kerman and 
Klein \cite{kk1,kk2,kk3,kk4}.  This theory starts from the spherical
shell model with residual two-body interactions.  A formally exact set of
equations is derived that
{\it resembles} the equations of a particle-core coupling model, but requires
even after a suitable restriction on the size of the configuration space,
the solution of a formidable non-linear problem.  In the KKDF model
the problem is linearized by choosing the matrix elements of multipole
and pairing operators, that arise from the interaction, either 
phenomenologically 
or from experiment.  If this can be done for a sufficiently extensive
model space, we can be reasonably confident that the corresponding
equations of the non-linear theory have been solved accurately (although
we have no proof of this statement).  The KKDF program can be carried out
at the present time only if the interaction chosen is sufficiently
simple that in fact the ``core'' matrix elements that enter can be 
determined at least in part from experiment.  

In previous work we have used a conventional Hamiltonian in 
which the interactions were confined to monopole pairing and quadrupole
multipole interactions.  Among the applications made, the simplest have been
to the low lying bands of several well-deformed odd rare earth nuclei.  To
a good approximation the only free parameter was the strength of the 
quadrupole interaction.  With the important exception to which this 
paper is devoted, we found good agreement with the energies
and electromagnetic moments (where measured) of all the lowest lying bands,
provided the calculations were carried out with three major shells for the
single-particle space and with the inclusion of excited bands of the core
(including but not confined to beta and gamma bands).
The inclusion of the excited
core bands was necessary to account for odd bands near the top
of the energy region considered, approximately 1 Mev.

The only discrepancy found was that our calculations predicted, uniformly, 
an exaggerated staggering of levels for $K=\frac{1}{2}$ bands compared 
to experiment.  In a particle-rotor model, it is easy to see from
standard formulas \cite{RS} that this discrepancy can be erased
by introducing the Coriolis attenuation factor.  As we have argued
above, the only fundamental way that is open to us to improve our 
calculations
is to add an interaction to the Hamiltonian. Such an interaction 
should effectively
contribute to our equations a Coriolis-like coupling that is  
dynamically independent of the standard one (the latter hidden in 
our formalism because we calculate in the laboratory system).
Once we have reached this point in our thinking,
it is almost obvious that the choice should be a magnetic dipole
interaction, also satisfying our requirement that the core matrix
elements can be measured. In fact, from the well-known properties of
vector operators it follows that contributions diagonal in 
both core angular momentum and particle angular momentum 
of any dipole interaction are exactly
equivalent to a Coriolis interaction.  However, the magnetic dipole
interaction can also have off-diagonal elements and therefore the equivalence
is only approximate.  

It may be noted that there is even a simpler way to obtain an additional
Coriolis coupling than the one we have chosen, and that is to add to the
Hamiltonian a term proportional to the square of the  angular momentum.
We shall verify that such an interaction can also be used to fit the
data.
Such a Coriolis coupling term was seen to arise, for example, 
in a paper \cite{FDSM} showing that a version of the 
particle-rotor model can be derived from the Fermion Dynamical Symmetry
Model (FDSM).  This is not surprising, since the standard Hamiltonian of 
FDSM indeed contains a term proportional to the square of the angular
momentum operator of that model.  This observation was not followed up.

It is also of interest to remark that in a recent analysis \cite{Zucker}
which sought to identify the most important multipole components of
a shell-model interaction, it was found that magnetic dipole interactions
were present.  The result was represented as a spin-spin interaction and
therefore different in detail from either of the possibilities examined
in this work.  Nevertheless this analysis further buttresses our viewpoint.

Finally we outline the contents of the body of work that follows. In 
Sec.\ II, we briefly review the basic elements of our theory and incorporate
the changes necessary to include the new interaction.  In Sec.\ III, we
specify how the parameters of the magnetic dipole interactions were chosen.
The results of our calculations are presented in Sec.\ IV, together with
a brief description of how the analysis was carried out, 
where it was deemed necessary
to supplement discussions in our earlier papers.   In Sec.\ V we compare
some results found for the magnetic dipole interaction with a corresponding
analysis in which we simply add to the Hamiltonian a term proportional
to the square of the angular momentum. A brief discussion of our results is
given in Sec.\ VI.

\section{Theory}
For the sake of completeness, we start with a brief review of the 
basic elements of our method.
We base our considerations on a shell-model Hamiltonian of 
the form,

\begin{eqnarray}
     H
   =
     \sum_{a}h_{a} a^{\dag}_{\alpha} a_{\alpha}
  &+&
       \frac{1}{4} \sum_{abcd} \sum_{LM_{L}}
         {F_{acdb}(L)}   \,
       {B^{\dag}_{LM_{L}}(a,c)} {B_{LM_{L}}(d,b)}
   \nonumber \\
   &+&
       \frac{1}{4}
       \sum_{abcd} \sum_{LM_{L}}
            G_{abcd}(L) \,
       A^{\dag}_{LM_{L}}(a,b) A_{LM_{L}}(c,d)
. \label{eq:m1}
\end{eqnarray}
Here $h_{a}$ ($\alpha=(j_{a},m_{a})$ and $a=j_{a}$)
are the spherical single particle
energies, $B_{LM}$  is the elementary particle-hole multipole operator,
\begin{equation}
       B^{\dag}_{LM_{L}}(a,b)
    \equiv
       \sum_{m_{a}m_{b}} s_{\beta} \;
       {C^{LM_{L}}_{\alpha\bar{\beta}}}  \;
       {a^{\dag}_{\alpha}} {a_{\beta}}
    =
       \left[
          a_{a}^{\dag} \times \tilde{a}_{b}
       \right] _{M_{L}}^{L} ,
\end{equation}
and  $A_{LM}$ is the elementary
particle-particle multipole operator

\begin{equation}
          A^{\dag}_{LM_{L}}(a,b)
       \equiv
         \sum_{m_{a}m_{b}}
          C^{LM_{L}}_{\alpha\beta}  \;
         {a^{\dag}_{\alpha}}{a^{\dag}_{\beta}}
       =
         \left[
            a_{a}^{\dag} \times a_{b}^{\dag}
         \right] _{M_{L}}^{L}
,
\end{equation}
where $C_{\alpha\beta}^{LM}$ are the Clebsch-Gordan coefficients
and $s_{\alpha}=(-)^{j_{a}-m_{a}}$.
The coefficients $F$  are the particle-hole matrix elements

\begin{equation}
    {F_{acdb}(L)}
    \equiv
       \sum_{m's}{s_{\gamma}}{s_{\beta}} \;
       {C^{LM_{L}}_{\alpha\bar{\gamma}}} \,
       {C^{LM_{L}}_{\delta\bar{\beta}}}     \,
   \,  V_{\alpha\beta\gamma\delta}
,
\label{eq:f}
\end{equation}
and $G$ the particle-particle matrix elements
\begin{equation}
     G_{abcd}(L)
   \equiv
      \sum_{m's}
     {C^{LM_{L}}_{\alpha\beta}}    \,
     {C^{LM_{L}}_{\gamma\delta}}   \,
     V_{\alpha\beta\gamma\delta}
.
\label{eq:g}
\end{equation}

The task is to obtain equations for the states and  
energies
of the odd nucleus assuming that the properties of neighboring even nuclei
are known.
The states of the odd nucleus (particle number $N$) are designated as
$\left| 
 \,J\mu\nu \right>$
where $\nu $ denotes all quantum numbers beside the angular
momentum $J$ and its projection $\mu$. The eigenstates and eigenvalues
of the neighboring even nuclei with particle numbers
($N\pm1$)
are $\left|IM\,n\,(N\pm1) \right>$ and
$E_{In}^{N\pm1}$, respectively,
where $n$ plays the same role for even nuclei as
$\nu $ does for the odd nuclei. The equations of motion (EOM)
are obtained by forming commutators between the Hamiltonian and single
fermion operators, leading to 
\begin{eqnarray}
     \left[ a_{\alpha} ,H  \right]
   =
     h_{a} a_{a}
  &+& \frac{1}{4} \sum_{bd \gamma}\sum_{LM}
     C_{\alpha \gamma}^{LM} G_{acbd}(L) a^{\dag}_{\gamma}A_{L M}(c,d)
   \nonumber \\
  &+& \frac{1}{4} \sum_{bd \gamma}\sum_{LM}
     {s_{\gamma}}C_{\alpha \bar{\gamma}}^{LM}
     F_{acdb}(L) a_{\gamma}B_{L M}(d,b)
, \label{eq:pqEOMa1}
\end{eqnarray}
together with its hermitian conjugate.

The matrix elements of these equations provide expressions that determine the 
single-particle coefficients of fractional parentage (cfp),
\begin{eqnarray}
V_{J\mu\nu}(\alpha; IMn)& =& \langle J\mu\nu|a_\alpha|IMn(A+1)\rangle , 
\label{eq:vee} \\
U_{J\mu\nu}(\alpha;IMn)&=& \langle J\mu\nu|a^{\dag}_{\bar{\alpha}}|
IMn(A-1)\rangle.  \label{eq:you}
\end{eqnarray}

In terms of a convenient and physically
meaningful set of energy differences and sets of multipole fields
and pairing fields defined below, we obtain generalized matrix
equations of the Hartree-Bogoliubov form   
\begin{eqnarray}
   {\cal E}_{J\nu}V_{J\mu\nu}(\alpha; IMn) 
&=&
   (\epsilon +\omega^{(A+1)} 
   +\Gamma^{(A+1)})_{\alpha IMn,\gamma I^{\prime} M' n'}
   V_{J\mu\nu} (\gamma;I' M' n')  
 \nonumber \\
&&      +\Delta_{\alpha IMn,\gamma I' M' n'}U_{J\mu\nu}(\gamma;I'M' n'), 
\label{eq:hfb1} 
\\
 {\cal E}_{J\nu}U_{J\mu\nu}(\alpha; IMn) &=&
  (-\epsilon +\omega^{(A-1)} 
 -\Gamma^{(A-1)\dag})_{\bar{\alpha} IMn,\bar{\gamma} 
I' M' n'}U_{J\mu\nu}(\gamma;I' M' n')   \nonumber \\
&&
 +\Delta^{\dag}_{\bar{\alpha} IMn,\bar{\gamma} I' M' n'}
V_{J\mu\nu}(\gamma;I'M' n'),  \label{eq:hfb2} 
\end{eqnarray}

where
\begin{eqnarray}
   {\cal E}_{J\nu} &=& -E_{J\nu} +\frac{1}{2}(E_0^{(A+1)}+E_0^{(A-1)}), 
\label{eq:def1}  \\
   \epsilon_{\alpha IMn,\gamma I'M'n'} &=& \delta_{\alpha\gamma}
      \delta_{II'}\delta_{MM'}\delta_{nn'}(h_a - \lambda_A), 
\label{eq:def2} \\
\lambda_A &=& \frac{1}{2}(E_0^{(A+1)}-E_0^{(A-1)}),  \label{eq:def3} \\
\omega^{(A\pm 1)}_{\alpha INn,\gamma I'M'n'} &=& \delta_{\alpha\gamma}
\delta_{II'}\delta_{MM'}\delta_{nn'}(E_{In}^{(A\pm 1)}-E_0^{(A\pm 1)}),
\label{eq:def4} \\
\Gamma^{(A\pm 1)}_{\alpha IMn,\gamma I'M'n'} &=& \frac{1}{2}\sum_{LM_L}
   \sum_{bd}s_\gamma C_{\alpha \bar{\gamma}}^{LM_L}\,F_{acdb}(L), \nonumber \\
&& \langle I'M'n'(A\pm 1)|B_{LM_L}(db)|IMn(A\pm 1)\rangle \label{eq:def5}\\
\Delta_{\alpha IMn,\gamma I'M'n'} &=&\frac{1}{2}\sum_{LM_L}\sum_{bd}
  C_{\alpha\bar{\gamma}}^{LM_L} \, G_{acdb}(L) \nonumber \\
&& \langle I'M'n'(A- 1)|A_{LM_L}(db)|IMn(A+1)\rangle. \label{eq:def6}
\end{eqnarray}
Here ${\cal E}$ is the eigenvalue, the energy of the state of an odd
nucleus relative to the average ground state energy of the neighboring 
even ones.  The remaining definitions refer to the elements of the effective
Hamiltonian matrix: 
$\epsilon$ are the single particle energies measured relative to
the chemical potential, $\lambda$ the chemical potential,
the elements of  $\omega$ are the 
excitation energies
in the even nuclei, and $\Gamma$ and $\Delta$ are the multipole and
pair fields, respectively. 

The solutions also require a normalization condition that is derived 
from matrix elements of the fundamental anticommutation relations,
\begin{equation}
    \frac{1}{\Omega} \sum_{\alpha \,IMn}
    \left[
         |U^{\nu }_{J\mu}(\alpha IM;n)|^2 + 
         |V^{\nu }_{J\mu}(\alpha IM; n)|^2
    \right]
        =1
,
\end{equation}
where
\begin{equation}
        \Omega = \sum_{j_{a}} (2j_{a} +1 ). 
\end{equation}

Since we shall use only a restricted set of interactions, specializing
to the case where $L=1,2$ multipoles and the $L=0$ pairing are included, 
the multipole and pairing fields defined above take the form
\begin{eqnarray}
\Gamma^{(A\pm 1)}_{\alpha IMn,\gamma I'M'n'} &=&
        \frac{1}{2}\sum_{M_1}   \sum_{bd}s_\gamma 
        ~C_{\alpha \bar{\gamma}}^{1M_1} F_{acdb}(1), \nonumber \\ \nonumber
       &&
    \langle I'M'n'(A\pm 1)|B_{1M_1}(db)|IMn(A\pm 1)\rangle \\ \nonumber
    && 
    + \frac{1}{2}\sum_{M_2}   \sum_{bd}s_\gamma 
        ~C_{\alpha \bar{\gamma}}^{2M_2} F_{acdb}(2), \nonumber \\
       &&
    \langle I'M'n'(A\pm 1)|B_{2M_L}(db)|IMn(A\pm 1)\rangle\\
  \Delta_{\alpha IMn,\gamma I'M'n'}   &=&
     \frac{1}{2}\sum_{bd} C_{\alpha\bar{\gamma}}^{00} G_{acdb}(0) \nonumber \\
&& \langle I'M'n'(A- 1)|A_{00}(db)|IMn(A+1)\rangle.
\end{eqnarray}
The further specialization to the magnetic dipole, 
the mass quadrupole, and the standard monopole pairing interactions, 
is defined by the equations
\begin{eqnarray}
  F_{acdb}(1) = -\kappa_1 \, M_{ab}(1) \, M_{dc}(1), \\ 
  F_{acdb}(2) = -\kappa_2 \, F_{ab}(2) \, F_{dc}(2),   \\
  G_{acdb}(0) = -g_0 \, G_{ab}(0) \, G_{dc}(0),  
\end{eqnarray}
together with the identification of the sums 
\begin{eqnarray}
{\cal M}_{\mu} = \sum_{ac} \, M_{ac}(1)  B_{1M_1}(a,c), \\ 
Q_{\mu} = \sum_{ac} \, F_{ac}(2)  B_{2M_2}(a,c), \\
\Delta = \sum_{ac} \, G_{ac}(0)  A_{00}(a,c), 
\end{eqnarray}
as the electromagnetic dipole, the mass quadrupole, and the pairing
gap operators, respectively.

With the definitions given above, the multipole and pairing fields
take the form
\begin{eqnarray}
\Gamma^{(A\pm 1)}_{\alpha IMn,\gamma I'M'n'} &=&
        -\frac{\kappa_1}{2}\sum_{M_1}  s_\gamma 
        ~C_{\alpha \bar{\gamma}}^{1M_1} M_{ac}(1) \nonumber \\ \nonumber
       &&
   \times \langle I'M'n'(A\pm 1)|{\cal M}_{M_1}(db)|IMn(A\pm 1)\rangle \\ \nonumber
    && 
    - \frac{\kappa_2}{2}\sum_{M_2}  s_\gamma 
        ~C_{\alpha \bar{\gamma}}^{2M_2} F_{ac}(2), \nonumber \\
       &&
 \times    \langle I'M'n'(A\pm 1)|Q_{M_2}(db)|IMn(A\pm 1)\rangle 
                \label{eq:Gamma} ,\\
  \Delta_{\alpha IMn,\gamma I'M'n'}   &=&
     -\frac{g_0}{2} C_{\alpha\bar{\gamma}}^{0} G_{ac}(0) \nonumber \\
&& \times \langle I'M'n'(A- 1)|\Delta|IMn(A+1)\rangle.
\end{eqnarray}


We apply the Wigner-Eckart theorem to obtain the equations of
motion in their final form.  For the multipole and pairing fields, we
use the following definitions of reduced matrix elements
\begin{eqnarray} 
  \langle I'M'n'(A\pm 1)|{\cal M}_{M_1}|IMn(A\pm 1)\rangle &=&  
          \frac{(-1)^{I-M}}{\sqrt{3}} \, 
          C_{IM \,I'-M'}^{1M_1} ~ \langle I'n'(A\pm 1)||{\cal M}||In
          (A\pm 1)\rangle , \\
  \langle I'M'n'(A\pm 1)|Q_{M_2}|IMn(A\pm 1)\rangle &=&  
          \frac{(-1)^{I-M}}{\sqrt{5}} \, 
          C_{IM \,I'-M'}^{2M_2} ~ \langle I'n'(A\pm 1)||Q||In(A\pm 1)\rangle , \\
     \langle I'M'n'(A\pm 1)|\Delta|IMn(A\pm 1)\rangle &=&  
           (-1)^{I-M} \, 
           C_{IM \,I'-M'}^{00}   \langle I'n'(A\pm 1)||\Delta||
           In(A\pm 1)\rangle ,
\end{eqnarray}
and for the cfp $V$ and $U$, we employ the definitions
\begin{eqnarray}
  V_{J\mu\nu}(\alpha,IMK) &=& \frac{(-1)^{J-\mu}}{\sqrt{2j_a +1}} \, 
              C_{IM\,J-\mu}^{\alpha}   ~ {\cal V}_{J\nu}(aIK),  \label{eq:WE1} \\
U_{J\mu\nu}(\alpha,IMK)&=&\frac{(-1)^{J-\mu +j_a +m_a}}{\sqrt{2j_a +1}}
 C_{IM\,J-\mu}^{\alpha} ~ {\cal U}_{J\nu}(aIK).  \label{eq:WE2} 
\end{eqnarray}

We thus find the equations of motion
\begin{eqnarray}
{\cal E}_{J\nu}~{\cal V}_{J\nu}(aIK)
&=&
        (\epsilon_a +\omega_{IK})~ {\cal V}_{J\nu}(aIK)
\nonumber \\
&& 
          +\sum_{cI'K'} \Gamma_{aIK,cI'K'}^{(A+1)}~{\cal V}_{J\nu}(cI'K') \nonumber \\
&&         +\sum_{cI'K'} \Delta_{aIK,cI'K'}~{\cal U}_{J\nu}(cI'K'), 
\label{eq:red1} \\
{\cal E}_{J\nu}~ {\cal U}_{J\nu}(aIK) 
&=& 
           (-\epsilon_a +\omega_{IK})~{\cal U}_{J\nu}(aIK)
\nonumber \\
&&    -\sum_{cI'K'} \Gamma_{aIK,cI'K'}^{(A-1)}~{\cal U}_{J\nu}(cI'K') \nonumber \\
&&    +\sum_{cI'K'} \Delta_{aIK,cI'K'}~{\cal V}_{J\nu}(cI'K'), 
\label{eq:red2} 
\end{eqnarray}

where

\begin{eqnarray}
 \Gamma^{(A\pm 1)}_{a IK,cI'K'} 
        &=& 
                - \frac{\kappa_1}{2} (-1)^{j_{c}+I+J}
        \left\{\begin{array}{ccc}
         j_{a} & j_{c} &  1 \\
         I'    & I     &  J 
        \end{array}\right\}
         ~ \langle I'K'(A\pm 1)||{\cal M}||IK(A\pm 1)\rangle  
          M_{ac}(1)                 \nonumber \\ 
          &-& 
                  \frac{\kappa_2}{2} (-1)^{j_{c}+I+J}
         \left\{\begin{array}{ccc}
         j_{a} & j_{c} &  2 \\
         I'    & I     &  J 
        \end{array}\right\}
         ~ \langle I'K'(A\pm 1)|| Q||IK(A\pm 1)\rangle 
          F_{ac}(2),                                                   \\
    \Delta_{aIn,cI'n'}        &=&  -\frac{g_0}{2}  
                \langle I'n'(A\pm 1)|| \Delta||In(A\pm 1)\rangle. 
\end{eqnarray}

\section{Determination of magnetic parameters}
In this section we describe how to extract the magnetic parameters
for the core and for the single particle motion.

\subsection{Phenomenology of the even core}
The matrix elements of the operator ${\cal M}$  between states of the 
even nuclei have to be determined. 
The reduced matrix elements of this operator  can be expressed 
using the Wigner-Eckart theorem as
\begin{equation}
   \langle I || {\cal M} || I \rangle  
   = \frac{ \sqrt{ 2I+1}}{C_{II10}^{II} } ~ \langle II |{\cal M}_0  | 
   II \rangle 
\end{equation}
From the definition of the intrinsic magnetic moment $\mu$ we have
\begin{equation}
  \mu= \langle II | {\cal M}_0 | II \rangle 
\end{equation}

Consequently we have 
\begin{eqnarray}
    \langle I || {\cal M} || I \rangle  & = &
         \frac{ \sqrt{ 2I+1}}{C_{II10}^{II} }~ \mu \\
\end{eqnarray}

Because of the limitation on available experimental data
for the intrinsic magnetic moments of the states of the neighboring 
even nuclei, we have chosen to represent their 
values by the simplest possible phenomenological model, 
\begin{equation}
  \mu=g_{R}(I) I \simeq g_{1} I,
\end{equation}
where the parameter $g_{R}$ is the effective $g$ factor for the 
rotational motion. Though this quantity can itself depend on the angular 
momentum, we have found that the use of a constant value, $g_1$, provides a 
reasonable, though not perfect, fit to the data and, at the same time, 
simplifies our study of the odd systems.
For the resulting values of $g_{1}$ see Table~\ref{tab:res1}.

\subsection{Matrix elements of the M1 operator for a single particle}
For the matrix elements of the M1 operator between single particle
states we quote the formula from \cite{RS}   
as 
 \begin{eqnarray} 
  \left< j_{a} \parallel {\cal M} \parallel j_{c} \right>&=&
          \left<n_{a}l_{a} j_{a} \mid  n_{c} l_{c} j_{c} \right>
           \sqrt{\frac{3\,(2j_{a}+1)\,(2j_{c}+1)}{4\pi}}
         (-1)^{j_{c}-1/2}
      \left(
         \begin{array}{ccc}
          j_{a} & 1 & j_{c} \\
          -1/2 & 0 &1/2
         \end{array}
      \right) \nonumber \\ 
        & & \hspace{2in} \times (1-k) \left[ \frac{1}{2}g_{s} 
-g_{l}(1+\frac{k}{2})\right]
, \label{eq:thm1}
\end{eqnarray}
where the first factor is a radial overlap integral, and
\begin{equation} 
        k = (j_{a} + \frac{1}{2})(-1)^{(j_{a} +l_{a} +1/2)} + 
         (j_{c} + \frac{1}{2})(-1)^{(j_{c} +l_{c} +1/2)}
.
\end{equation}
Here $g_{s}$ and $g_l$ are the spin and orbital gyromagnetic
ratios. In our calculations, we have adopted the values suggested by the 
analysis found in \cite{BM}, namely 0.70 of their bare values.

\section{Applications}
We have applied the theory described above to four nuclei, the odd neutron
nuclei $^{157}$Gd, $^{159}$Dy, $^{163}$Er and the odd proton case
 $^{157}$Tb.  All but $^{163}$Er were studied earlier \cite{pav:2,pav:5} 
by our methods.   The results for the neutron cases, in particular, 
displayed $K=\frac{1}{2}$ staggering not seen in the experimental
results.  We added Er to the mix because it has several $K=\frac{1}{2}$
bands and has been used in the past 
as a classic example of the Coriolis attenuation phenomenon.

The special technical problems involved in applying our formalism have 
been fully documented and solved in our previous papers and will
not be repeated here.  As far as the size of the model space is concerned
we remind the reader that for the low energy data that we are analyzing, 
it suffices to describe the core by its ground-state band and those few 
excited bands, such as beta and gamma bands (but not restricted to these)
that show some quadrupole collectivity with respect to the ground band.   
We then choose the size
of the single-particle space coupled to these bands large enough so that
further enlargement does not modify the results.  In this respect we can
claim to have essentially exact solutions of our semi-microscopic model,
though the question of whether these are close to fully self-consistent
solutions of the Kerman-Klein equations remains an open question.  In general
we have found that three major shells centered at the valence shell suffices.

For all intents and purposes, our previous calculations contained only one
free parameter, the strength $\kappa_2$ of the quadrupole-quadrupole
interaction, all other parameters having been determined by other
experimental results.  (A minor cavil to this assertion is that small
adjustments of the spherical single-particle energies emerging from
Woods-Saxon calculations were permitted for levels near the Fermi surface.)
In the present work, we have a second parameter, the strength $\kappa_1$
of the magnetic interaction, to adjust.  

\begin{figure}[h]
  \begin{center}
    \leavevmode
\centerline{\hbox{\psfig{figure=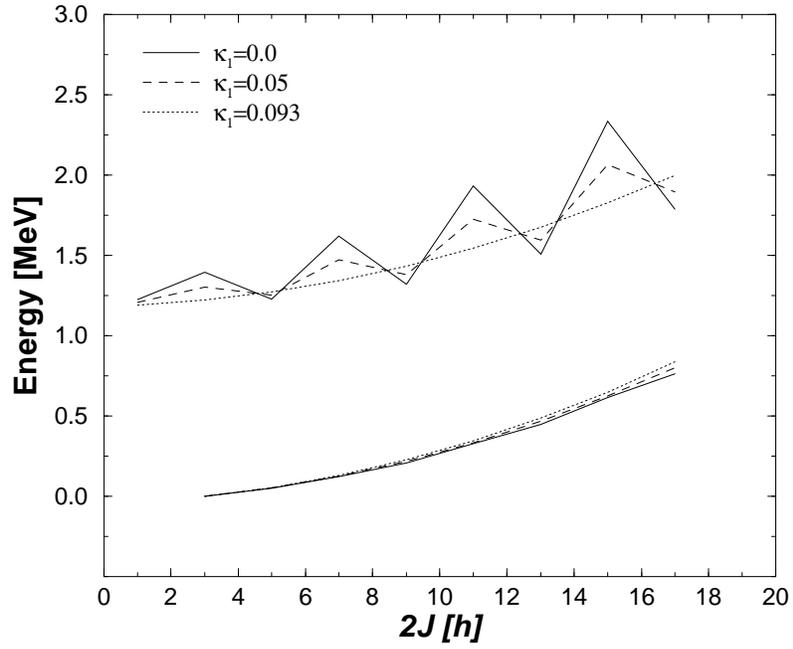,height=10cm}}    }
  \end{center}
  \caption{A model calculation to illustrate the effect of the magnetic 
dipole-dipole interaction on staggering.
 A single-particle space with a single $j$ value is coupled 
to the ground state bands of the even neighbors of $^{157}$Gd. The strength
of the quadrupole force is fixed at $\kappa_{2}=0.564 \rm{MeV/fm}^{2}$. 
The strength
of the magnetic dipole force is varied to demonstrate the effect of this
interaction.  } 
  \label{fig:K1effect}
\end{figure}

\begin{figure}[h]
  \begin{center}
    \leavevmode
\centerline{\hbox{\psfig{figure=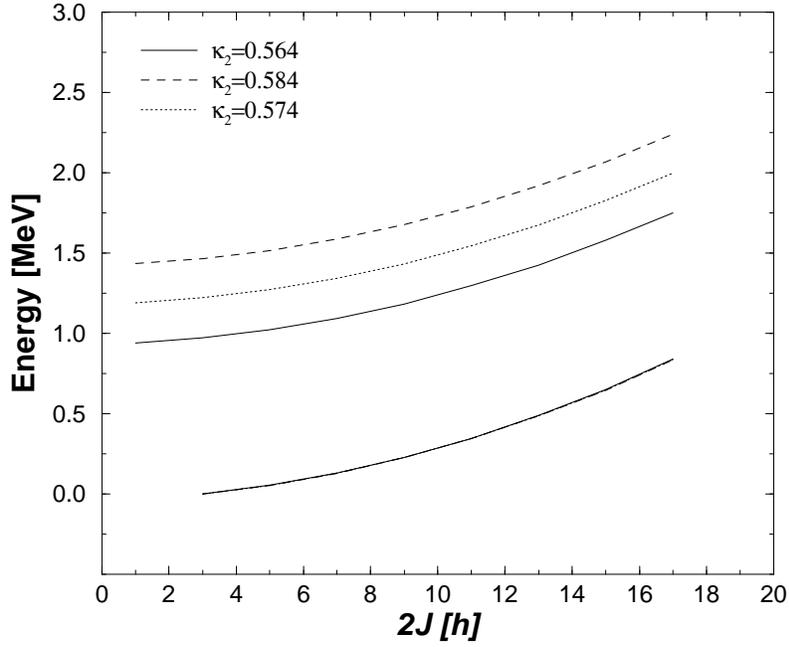,height=10cm}}    }
  \end{center}
  \caption{The purpose of this figure is to demonstrate that the strength 
of the quadrupole
field does not influence the staggering. The model is the 
same as in Fig~\protect{\ref{fig:K1effect}}. The value, $\kappa_1$, of the magnetic 
interaction is set at the value that minimizes
the staggering of the $K=1/2$ band. Then the effect of varying
$\kappa_2$ is studied.} 
  \label{fig:K2effect}
\end{figure}

Before displaying the results
found using the full theory, we illustrate the
method of fitting the data with the additional parameter by studying a 
simplified model involving the coupling of a single $j$ level to the 
core bands of Gd and with a value of the 
quadrupole interaction chosen so that we have a lower $K=\frac{3}{2}$ 
band clearly separated from a higher $K=\frac{1}{2}$ band.  We see from 
Fig.~\ref{fig:K1effect}
that for fixed $\kappa_2$, the staggering  is reduced and finally eliminated
as we raise $\kappa_1$ from its initial value of 0.  At the same 
time the effect on the shape of the $K=\frac{3}{2}$ band is negligible.   
Next, fixing the magnetic interaction at its optimum value, we see from 
Fig.~\ref{fig:K2effect}
that changing the value of $\kappa_2$ shifts band-head energies.  What
is not clear from these figures is that in the general case, the magnetic
interaction also shifts relative band-head energies.  Therefore in actual
calculations, we first adjust $\kappa_2$ to give the best results for
other than the staggering, next find the optimum value of $\kappa_1$, and
subsequently refit band heads with a new search on $\kappa_2$.  This is
an iterative procedure to find the best pair of coupling strengths.

\begin{figure}[h]
  \begin{center}
    \leavevmode
\centerline{\hbox{\psfig{figure=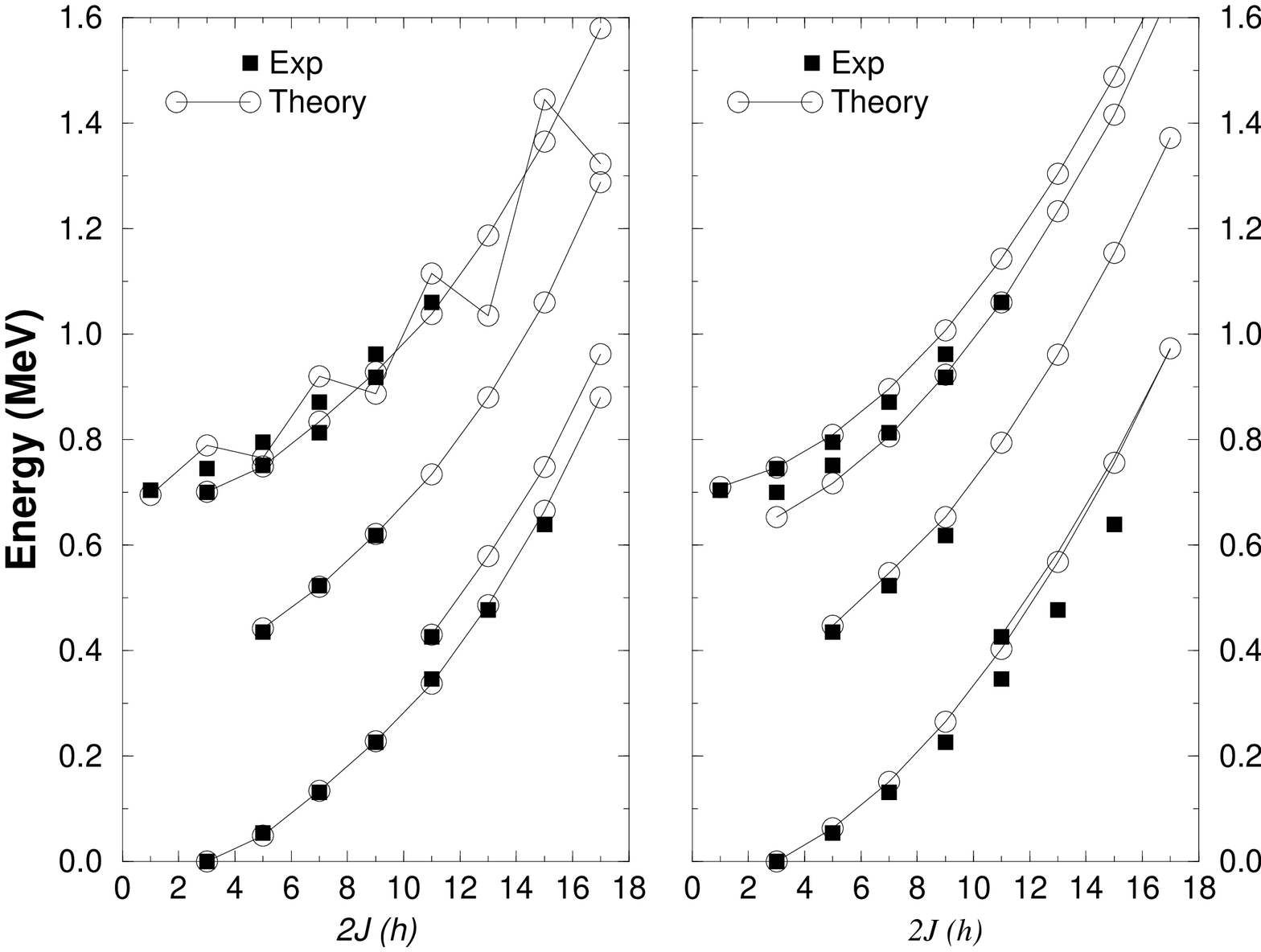,height=10cm}}    }
  \end{center}
  \caption{Negative parity energy levels of $^{157}$Gd. The calculation
includes single particle
states for three major shells and  excited bands of the
core. On the left is the previous calculation, with the 
strength of the quadrupole force set at
0.380 MeV/fm$^{2}$ and $\kappa_1 =0.0$ MeV/$\mu_{\nu}^{2}$. On the right 
is the improved calculation with
 $\kappa_2=0.394$ MeV/fm$^{2}$ and $\kappa_1 =0.045$
MeV/$\mu_{\nu}^{2}$. }
  \label{fig:Gd157}
\end{figure}

\begin{figure}[h]
  \begin{center}
    \leavevmode
\centerline{\hbox{\psfig{figure=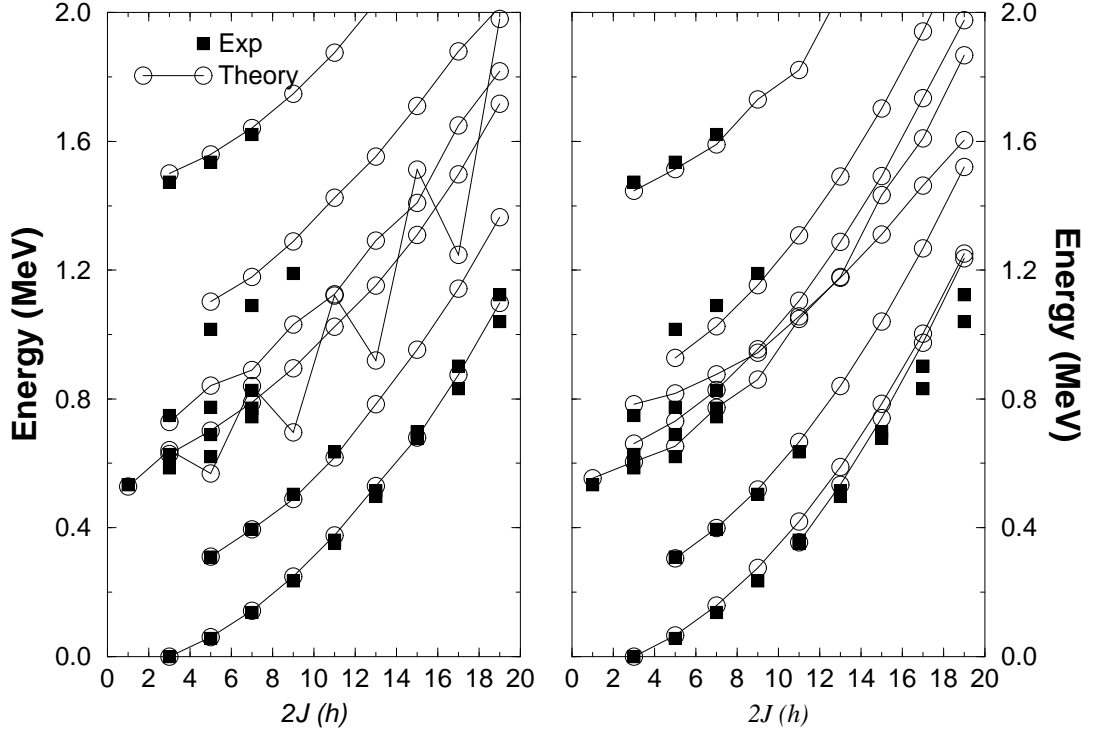,height=10cm}}    }
  \end{center}
  \caption{Negative parity energy levels of $^{159}$Dy. The calculation
includes single particle
states for three major shells and excited bands of the
core. On the left is the previous result with the strength of the 
quadrupole force set at 
0.383 MeV/fm$^{2}$ and $\kappa_1 =0.0$ MeV/$\mu_{\nu}^{2}$. 
On the right is the improved calculation with $\kappa_2=0.368$ MeV/fm$^{2}$ 
and $\kappa_1 =0.043$ MeV/$\mu_{\nu}^{2}$. }
  \label{fig:Dy159}
\end{figure}

We next present the results of our calculations.  The odd neutron nuclei
$^{157}$Gd and $^{159}$Dy were studied previously in \cite{pav:2}.
We found that we could fit the energies satisfactorily except
that a staggering of a $K=1/2$ band for each nucleus that was not observed
experimentally.  To obtain the quality of agreement found in the previous
work, it was necessary to include
the $\beta$ band, the $\gamma$ band, and an additional $K=0$ 
excited band.
In Figs.~\ref{fig:Gd157} and ~\ref{fig:Dy159} we show the
results of the augmented calculations and compare them to those found
previously.
It can be seen that the addition of the magnetic dipole interaction
removes the unphysical staggering found earlier without any serious
damage to the rest of the fit.  

\begin{figure}[h]
  \begin{center}
    \leavevmode
\centerline{\hbox{\psfig{figure=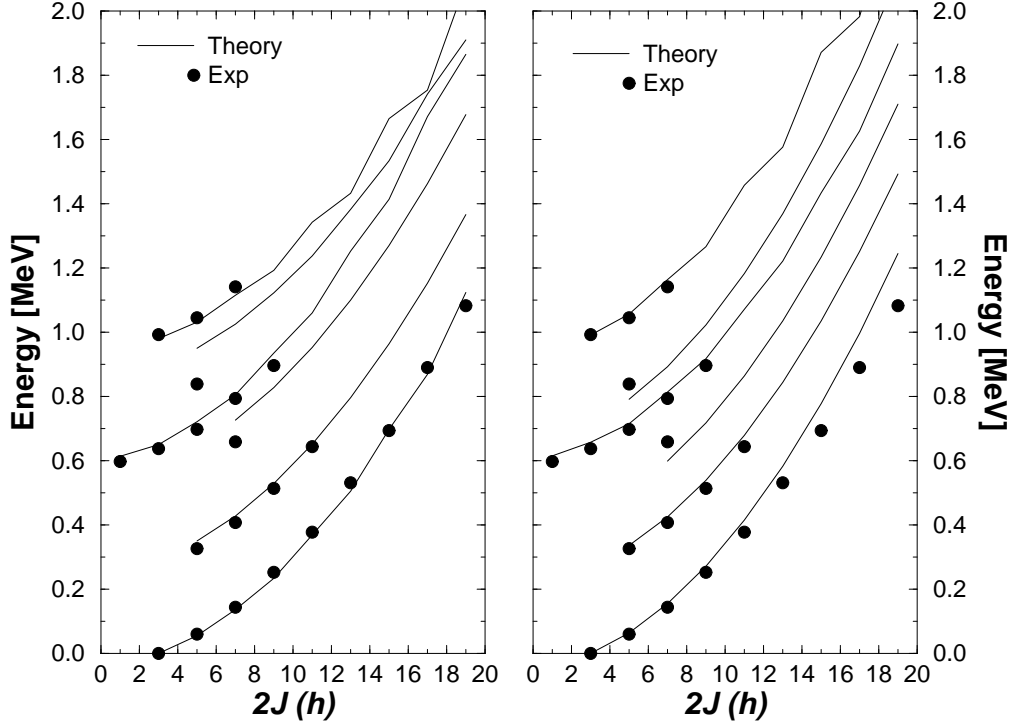,height=10cm}}    }
  \end{center}
  \caption{Positive parity energy levels of $^{157}$Tb. The calculation
includes single particle
states for three major shells and excited bands of the
core. On the left is shown the previous result where the strength 
of the quadrupole force is
0.428 MeV/fm$^{2}$ and $\kappa_1 =0.0$ MeV/$\mu_{\nu}^{2}$.
On the right is the improved calculation with $\kappa_2=0.420$ MeV/fm$^{2}$ 
and $\kappa_1 =-0.052$ MeV/$\mu_{\nu}^{2}$. }
  \label{fig:Tb157}
\end{figure}

The odd proton nucleus
$^{157}$Tb  was used in a previous study of the status 
of the traditional core-particle model as an approximation to the KKDF model
\cite{pav:5}.  In Fig.~\ref{fig:Tb157} we see that in this case
no exaggerated staggering was predicted by the theory without a magnetic 
dipole interaction.  The addition
of a magnetic dipole-dipole field with strength of the same order of
magnitude as for the other two nuclei gives, nevertheless, 
a perceptibly improved overall fit to the data. 

\begin{figure}[h]
  \begin{center}
    \leavevmode
\centerline{\hbox{\psfig{figure=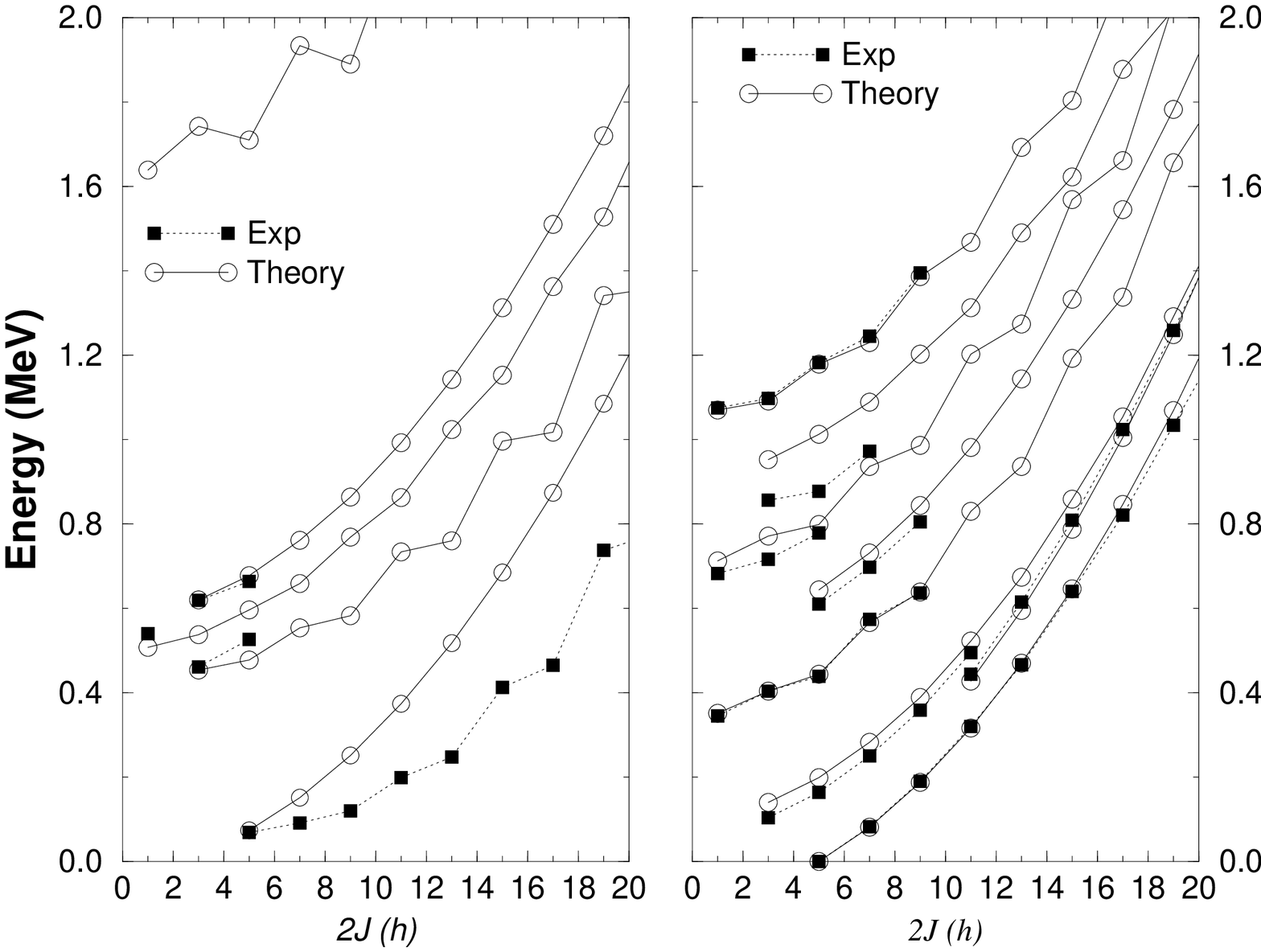,height=10cm}}    }
  \end{center}
  \caption{Energy levels of $^{163}$Er. Single particle
states for three major shells and excited bands of the
core are included.   On the left are the positive parity states
and on the right the negative parity states. For the positive parity
states the strength of the quadrupole force is 0.413 MeV/fm$^{2}$ when
$\kappa_1 =0.0$ MeV/$\mu_{\nu}^{2}$ (not shown) and 
$\kappa_2=0.399$ MeV/fm$^{2}$
when $\kappa_1 =-0.040$ MeV/$\mu_{\nu}^{2}$. For the negative parity
states the strength of the quadrupole force is 0.375 MeV/fm$^{2}$ when
$\kappa_1 =0.0$ MeV/$\mu_{\nu}^{2}$ (not shown) and 
 $\kappa_2=0.366$ MeV/fm$^{2}$
when $\kappa_1 =-0.040$ MeV/$\mu_{\nu}^{2}$. }
  \label{fig:Er163}
\end{figure}

Last we have applied our theory to the case of $^{163}$Er. We have
studied both the negative and positive parity states of this nucleus.
The results are shown in Fig.\ ~\ref{fig:Er163}.  In the right panel,
which displays the results for the levels of odd parity,
one sees that we have found a good fit to the data, which includes
three $K=\frac{1}{2}$ bands.  Remark that the fit is best for those two
of these bands for which the data is more extensive.  In the left panel,
which is devoted to the positive parity levels, there is only a single
$K=\frac{1}{2}$ bandhead.  The lowest band, $K=\frac{5}{2}$, the only
one that seems to be well-established experimentally, has features that
we have failed to capture thus far.

\section{{\bf J}$\cdot${\bf J} interaction}
\begin{figure}[h]
  \begin{center}
    \leavevmode
\centerline{\hbox{\psfig{figure=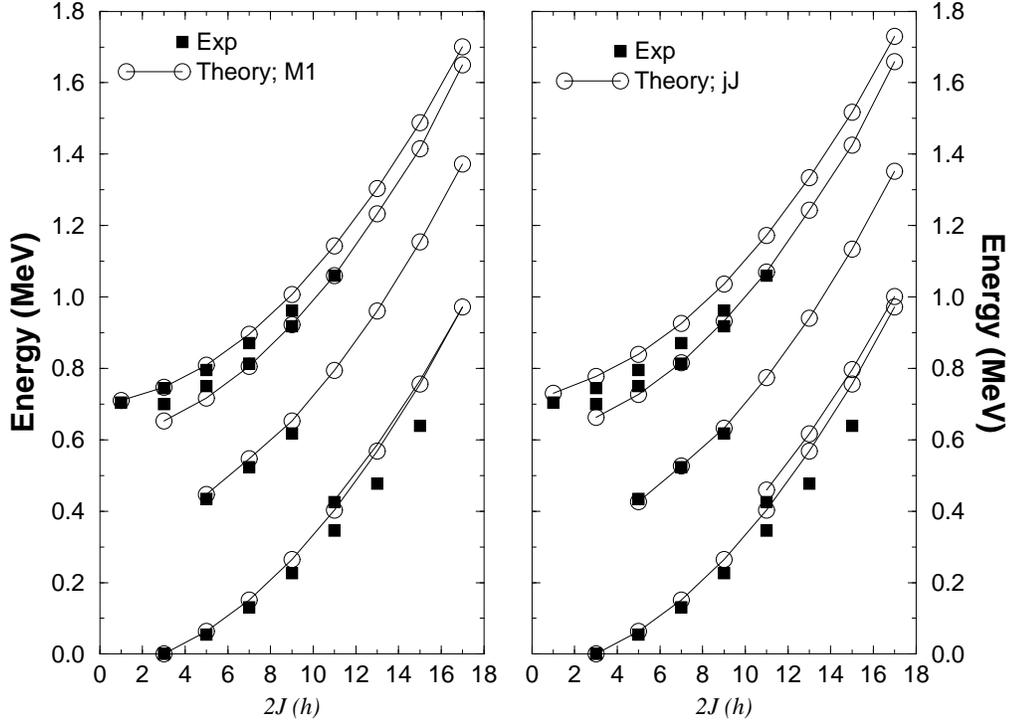,height=10cm}}    }
  \end{center}
  \caption{Comparison of two magnetic interactions.
On the left is the same fit as in Fig.~ \protect{\ref{fig:Gd157}} using the 
magnetic dipole operator and on the right the fit using the angular
momentum form.  For the calculations in the left panel, we 
have $\kappa_2 = 0.394$ MeV/fm$^2$ with $\kappa_1=0.045$ MeV/$\mu_\nu^2$;.
for the calculation on the right, $\kappa_2=0.392$ MeV/fm$^2$ and 
 $\kappa'_1=0.026$ MeV/$\mu_\nu^2$. }
  \label{fig:JJ}
\end{figure}

As mentioned in the introduction we can achieve results similar
to those reported above by replacing the magnetic dipole interaction 
used previously by a term
proportional to the square of the angular momentum.   In Fig.\ 
~\ref{fig:JJ} we contrast the new calculation thus engendered for
$^{157}$Gd (right panel) with that found in the previous section
(reproduced in the left panel.  We find that both interactions have a similar
effect on the staggering, but that the effect on the relative bandhead energies
is different.  In general, fits of comparable quality are obtained.
To achieve the desired  staggering, the strength, $\kappa'_1$ 
of the {\bf J}$\cdot${\bf J}
interaction should be set at $\kappa_1=0.026$.
We get the best fit if the 
strength of the quadrupole field $\kappa_2$ is altered slightly 
from what we had before: $\kappa_2=0.392$ MeV/fm$^{2}$ compared
to $\kappa_2=0.394$  MeV/fm$^{2}$.

We compare the two interactions to understand their similarities and
differences.  If only matrix elements diagonal in both the core and 
the particle are considered, it is a well-known property of vector
operators that the two interactions are equivalent.  In fact this
equivalence holds very well for the core, since for $K=0$ bands,
with $I$ values differing by two, the magnetic dipole operator has
no off-diagonal elements.  It is true that we lose this simplification
for the gamma band, but the physics we are discussing is dominated
by the ground band. When we turn, however, to the single-particle
space, off-diagonal matrix elements of the magnetic moment operator can 
play an important role, as in generating magnetic dipole transitions, and 
are thus the agent responsible for the slightly different results found
for the two interactions considered.

\section{Discussion}

The aim of this paper has been to present a reasonably convincing
case, both on the basis of \`a priori arguments and by illustrative 
calculations, that we have uncovered the origin of the phenomena
referred to as attenuation of the Coriolis coupling.  By basing
our considerations on a theoretical framework of great generality, we were
able to rule out a large subset of explanations offered previously, since
we obtained evidence of the attenuation phenomenon even when the physics
associated with these explanations was naturally incorporated in our
calculations.  The only possibility remaining to us was to think
about adding some term to the pairing and quadrupole interactions that
define the Hamiltonian used in our previous work.  Because of the 
nature of the effect sought, the most obvious choice is a magnetic
dipole interaction.  There are two such interactions that can be fitted
easily into our semi-microscopic theory, the scalar product of the 
magnetic moment operator with itself and the square of the angular
momentum operator.  Calculations done with the former on four
different nuclei showed that the excessive staggering encountered in
previous calculations, but absent experimentally, could be made to
disappear with a dipole interaction strength that remains constant within
10\% for the nuclei studied. 
For one of the nuclei, $^{157}$Tb, where no excessive staggering
was calculated previously, we found a slight improvement to the overall
fit with the same interaction strength.

For one case, that of $^{157}$Gd, we compared the results found for the two
types of interaction mentioned above and found quite similar, but not 
identical, results.   This suggests that we have identified the multipolarity
of the additional interaction necessary to explain the quenching effect,
but not its detailed form.

\section*{Acknowledgment}

We are grateful to Jolie Cizewski for a remark that stimulated this research.
Our work was supported in part by the U.\ S.\ 
Department of Energy under Grant No. 40264-5-25351


\begin{table}
\begin{center}
        \begin{tabular}{||c|c|c|c||}
        nucleus & $g_{1}$ [$\mu_{\nu}$]&  $\kappa_{1}$  [MeV/$\mu^2_{\nu}$]& $\kappa_{2}$   [MeV/fm$^2$] \\
        \hline \hline 
        $^{157}$ Gd & 0.371 &  0.045 & 0.394  \\ \hline 
        $^{159}$ Dy & 0.345 &  0.043 & 0.368  \\ \hline               
        $^{157}$ Tb & 0.362 &  0.052 & 0.420  \\ \hline               
        $^{163}$ Er (-) & 0.353 &  0.04 & 0.366  \\ \hline            
        $^{163}$ Er (+) & 0.353 &  0.04 & 0.396  \\
        \end{tabular}           
\end{center}
\caption{Parameters used in the calculations reported in this work. The 
second column lists the gyromagnetic ratios of the core nuclei, the
third column the values of the magnetic dipole interaction strenght,
and the last the values for the quadrupole interaction strenght. }
\label{tab:res1}

\end{table}

\end{document}